\newcommand{\beq}{\begin{equation}}
\newcommand{\eeq}{\end{equation}}
\newcommand{\bea}{\begin{eqnarray}}
\newcommand{\eea}{\end{eqnarray}}
\newcommand{\bear}{\begin{eqnarray*}}
\newcommand{\eear}{\end{eqnarray*}}

\newcommand{\rf}[1]{(\ref{#1})}
\newcommand{\bbb}{\begin{equation}}
\newcommand{\fff}{\end{equation}}

\documentclass[twocolumn,showpacs,preprintnumbers,amsmath,amssymb]{revtex4}
\usepackage{graphicx}% Include figure files
\usepackage{dcolumn}% Align table columns on decimal point
\usepackage{bm}% bold math  

\begin{document}

\draft

\title
{ANOMALOUS TAG DIFFUSION  IN THE ASYMMETRIC EXCLUSION MODEL 
 WITH 
PARTICLES OF ARBITRARY SIZES}

\author{
Anderson A. Ferreira}     
\affiliation{
Departamento de F\'{\i}sica, 
Universidade Federal de S\~ao Carlos, 13565-905, S\~ao Carlos, SP
Brazil}
\author{Francisco C. Alcaraz}
\email{alcaraz@if.sc.usp.br}
\affiliation{
Universidade de S\~ao Paulo, Instituto de F\'{\i}sica de S\~ao Carlos,
C.P. 369,13560-590, S\~ao Carlos, SP, Brazil}

\begin{abstract}

Anomalous behavior of correlation functions of tagged particles are 
studied in generalizations of the one dimensional asymmetric 
exclusion problem. In these generalized models the range of the 
hard-core interactions are changed and the restriction of 
relative ordering of the particles is partially brocken. The models 
probing these effects are those of biased diffusion of 
particles having size $S=0,1,2,\ldots$, or  an effective 
negative "size" $S=-1,-2,\ldots$, in units of lattice space. 
Our numerical simulations show that irrespective of the range of 
the hard-core potential, as long some relative ordering of 
particles are kept, we find suitable sliding-tag correlation 
functions whose fluctuations growth with time anomalously slow 
($t^{\frac{1}{3}}$), when compared with the normal diffusive 
behavior ($t^{\frac{1}{2}}$). These results indicate that the 
critical behavior of these stochastic models are in the 
Kardar-Parisi-Zhang (KPZ) universality class. Moreover a previous Bethe-ansatz 
calculation of the dynamical critical exponent $z$, for size $S \geq 
0$ particles is extended to the case $S<0$ and the KPZ result 
$z=\frac{3}{2}$ is predicted for all values of $S \in {Z}$.

\end{abstract}


\maketitle
\narrowtext    

	Dynamic and equilibrium properties of lattice gases, i. e., 
systems involving randomly moving particles with hard-core 
interactions, have received much attention within the last 
decades, mainly due to their intrinsic and nontrivial many 
body behavior. Several important results have been 
obtained for such systems of particles. In particular 
the study of tagged-particle diffusion in hard-core lattice gases 
with nearest-neighbor hopping revealed interesting nontrivial 
many body behavior \cite{harris}-\cite{binder}. 
In one dimension the rate of 
diffusion depends strongly if the diffusion is symmetric or 
not, i. e., has or not a prefered direction. In the symmetric 
case, the root-mean-squared displacement of a tagged 
particle increases as $t^{\frac{1}{4}}$ during a time interval $t$, 
  behavior that is anomalously slow when 
compared with the asymmetric case where this quantity exhibits 
the standard diffusive behavior $t^{\frac{1}{2}}$.  
Majumdar and Barma \cite{maba} have shown that even in the 
asymmetric case it is possible to find anomalous behavior of 
special correlations of tag particles (sliding-tag correlations), 
where the difference 
of the tag coordinates of the particles changes in time with 
a fixed tag velovity $v_{tag}$. For arbitrary values of 
$v_{tag}$, except for $v_{tag} = v_{tag}^{c}$, these correlations 
increase with a typical diffusive behavior $t^{\frac{1}{2}}$, 
while at the special value $v_{tag}=v_{tag}^c$, that 
depends on the density of particles, they increase 
anomalously as $t^{\frac{1}{3}}$. This last behavior can be 
understood by mapping the asymmetric diffusion model into 
the particle height interface model, whose fluctuations in the 
continuum limit are governed by the Kardar, Parisi and Zhang (KPZ) 
model \cite{kpz}.

From the above results \cite{maba} it is clear that the 
relevant parameters ensuring anomalous behavior for 
correlations of tag particles are:  the hard-core effect (that 
 implies, due to the topology of a one dimensional lattice, 
 a fixed relative ordering of particles), the right-left 
asymmetry of the diffusion, and the velocity in the tag space 
where the correlations are measured.

All these results are obtained by imposing that the tag 
particles has a unity size, in terms of 
lattice units, or equivalently the particles have 
a fixed relative order and two particles at the same site have 
an infinite hard-core repulsion.  The aim of this paper is to
obtain the anomalous behavior of these sliding-tag correlations 
in the case where we increase or decrease the  range of exclusion 
of molecules as well we relax the constraint of relative ordering 
of the molecules. 
The variation of the exclusion effect will be done by considering 
 the asymmetric diffusion of particles with arbitrary integer 
sizes 
$S=0,1,2,\dots$, and the effect of ordering will be verifyed 
by considering a generalization of the model where two 
consecutive ordered particles, can violate partially their 
ordering. This last model, as we shall see, corresponds 
to a generalization to the case where the particles in the 
asymmetric diffusion have an effective negative "size" 
$S=-1,-2,\dots$.

	We consider initially a one dimensional periodic lattice 
of  $N_s$ sites contaning $N_p$ particles of 
a fixed size $S$ ($S=0,1,2,\ldots$). These particles occupy 
$S$ consecutive sites, having a hard-core repulsion of 
range $S$. In the case $S=0$ the range of infinite repulsion 
 is zero and we may put an arbitray number of particles in a 
given site.  The dynamics of the asymmetric diffusion 
is such that  a particle is selected at random and with 
probability $p$ ($q$) the particle attempts to move by 
one unity of lattice spacing to the right (left). The movement 
is accepted if the hard-exclusion allows the final 
configuration. The unit of time is given by $N_p$ attempts 
of motion.  For a given number of particles $N_p$ the number 
of holes (positions where the particles are allowed to move) 
is given by $N_s -SN_p$ and the mean velocity $v_p$ of the 
particles is given by 
\begin{equation} \label{eq1}
v_p = \frac{(p-q)(N_s -SN_p)}{N_p +N_s -SN_p} = (p-q)(1-\chi),
\end{equation}
where $\chi = N_p/(N_s -(S-1)N_p)$ plays the role of an effective 
density of particles in the system. 
A given configuration at time t is given by the lattice 
coordinates of the particles $\{y(n,t)\}$, where 
$n=1,2,,\ldots,N_p$ are the sequential labels indexing the  particles. 
Since 
independently of the size of molecules the asymmetric diffusion 
process does not change their order (even for the size $S=0$ 
particles), we may chose this ordering  such that $y(n+1,t) 
\ge y(n,t) +S$, and the periodic boundary condition translate 
into $y(n\pm N_p,t) = y(n,t) \pm N_s$.
A corresponding interface fluctuation model can be defined by 
considering $y(n,t)$ as the local height of the surface at 
the horizontal cordinate $n$. In this model, at each time step,
with probability $p$ ($q$) $y(n,t)$ increase (or decrease) by 1, 
and the motion is performed if $y(n+1,t)-y(n,t) >S$ (or 
$y(n,t) -y(n-1,t) >S$). In the steady state, for $p>q$  (or 
$p<q$) the surface grows (or decreases) in the vertical 
direction, with velocity $v_p$.

In order to search for anomalous behavior of correlations we 
consider a set of correlations among the coordinates
$\{y(n_t,t)\}$, of the tagged particles with label 
$\{n_t\}$ at time $t$ and the cordinate $y(n_0,0)$ of a given 
particle $n_0$ at initial time $t=0$. The label difference of 
the tagged particles  
 $n_t - n_0 = v_{tag}t$ changes with time,
and  for convenience we 
write the tag velocity as $v_{tag} = \rho v_p b$, where 
 $\rho = N_p/N_s$ is the density of particles and 
$b = b(S,\chi)$ 
is a real parameter. The mean values of these coordinates 
\bea \label{eq2}
<y(n_t,t)-y(n_0,0)> &=& <y(n_t,0)-y(n_0,0> + v_p t \nonumber \\
&=& 
(v_p +v_{tag}/ \rho )t=v_F t
\eea
increases with the frame velocity $v_F= v_p (1 +b)$. The 
sliding-tag correlation functions \cite{maba}-\cite{binder} 
we considered are given by
\beq \label{eq3}
\sigma(v_{tag},t)^2 = <[y(n_t,t) -y(n_0,0)]^2> - v_F^2 t^2.
\end{equation}
We have measured these correlations by extensive Monte Carlo 
simulations by varying the asymmetry parameters $p,q$, 
the density of particles $\rho$, and the size $S$ of the 
particles. We verified that irrespective of the size $S$ of 
the particles, as long the diffusion is biased ($p \neq q$), 
anomalous behavior   
happens for special values of the tag velocity $v_{tag}$. 
 The value of $v_{tag}$ 
 depends on  the density $\rho$, the asymmetry parameter ($p-q$)
 and the 
size $S$ of the particles. Examples of these measurements, for the 
cases where $S=2$ and $S=0$ are shown respectively 
 in Fig. 1 and 2, where the 
time dependence of the correlation \rf{eq3} are shown for 
several values of the tag velocity. In Fig. 1 we consider 
$p =0.75$, $q =0.25$ and density $\rho=1/5$ in a lattice of 
$N_s =150\; 000 $
sites while in Fig. 2 $p =0.75$, $q=0.25$ and $\rho = 1/4$ 
in a lattice of $N_s =120\; 000$ sites. 
The average slope of the curves on these figures shows that apart 
from 
crossover effects, due to the  
finite size of the lattice, we have a 
normal diffusive behavior for arbitrary 
 values of $v_{tag}$, except 
at $v_{tag} = v_{tag}^c = b_c\rho v_p$, where 
$b_c =0.4167 $ for $S=2$ and $b_c = 0.2$ for $S=0$. 
At those critical values the increase of the correlations is 
anomalously slow and our numerical results indicates 
$\sigma \sim t^{\nu}$ where $\nu = 0.32 \pm 0.01$, for the $S=0$ 
particles 
and $\nu=0.33 \pm 0.01$ for the $S=2$ particles.

A constraint on the dependence of the parameter $b_c = b_c(S,\chi)$ 
can be derived by generalizing the arguments presented in \cite{maba} 
for 
unity size particles. Instead of describing the 
asymmetric diffusion in terms of the size $S$ particles we 
consider the diffusion of holes in the reverse order. This is 
possible since the holes does not change their relative order and 
behave as ordinary size $S=1$ particles 
traveling on an effective lattice with $N_s -N_p(S-1) = 1 -\chi(S-1)$ sites
(allowed positions).  
The mean velocity of holes is $v_h = S\chi (q-p)$, where the 
 factor $S$ appears because in their motion the holes jump particles 
with size 
$S$ \cite{obs1}. 
We can consider an equivalent sliding-tag 
correlation function (\ref{eq3}) for the holes, with the 
tag velocity $v_{tag}^h = (1-S\rho)b^h(\chi) v_h$ and the 
frame velocity $v_F^h = (1-b^h(\chi))v_h$. The correspondence 
between $b(S,\chi)$ and $b^h(\chi)$, describing the same 
fluctuations in both correlations (particles or holes) are 
obtained by imposing the equality \cite{maba} $v_F = v_F^h$. 
 At the critical point hole and particle correlations should 
reveal the same anomalous behavior. Taking into account that 
the holes behave as ordinary $S=1$ particles, but traveling 
$S$ times faster in an effective lattice with $N_s -(S-1)N_p$ 
sites, we have by comparing the tag velocities, 
$(N_s-(S-1)N_p)Sb_c^h(\chi) = N_sb_c(1,1-\chi)$. As a consequence 
we obtain the general relation

\begin{eqnarray} \label{eq4}
&&\chi (1-\chi (1-S))b_c(1,1-\chi) +(1-\chi)b_c(S,\chi) =  
\nonumber \\
&& 1 + (S-1)\chi, 
\end{eqnarray}
that recovers the known   relation  in the particular 
case where $S=1$ \cite{maba}. 

Our numerical experiments show  that 
  the critical values $b_c(S,\chi)$ do not 
depend on the asymmetry parameter as long the diffusion is 
asymmetric ($p \neq q$). Moreover our results are consistent 
with the conjecture
\begin{equation} \label{eq5}
b_c(S,\chi) =  (\frac{S}{1-\chi}+ 1 - S)\chi.
\end{equation}
The agreement of the above conjecture 
was tested exhaustively in numerical simulations for several  
 values  of $\rho$, $p$, $q$ and 
$S$. 
In Table I we present some of our estimates for the exponent 
$\nu$ at $p=0.75$ and using in the simulations the 
conjectured value $b_c(S,\chi)$ giving in (\ref{eq5}). 
Our results are clearly consistent with the exponent 
$\nu =\frac{1}{3}$ for all values of $S$ and $\chi$.

Up to now our results indicate that for arbitrary range of 
exclusion (size $S=0,1,2,\dots$) we can define appropriate 
sliding-tag correlation functions exhibiting anomalous 
behavior. The existence of such anomalous behavior is 
connected to the fact that  the dynamical behavior of these 
exclusion models are in 
the KPZ universality class. In fact   
 the Bethe-ansatz solution \cite{alcbar1} 
 of the  quantum Hamiltonian associated to the asymmetric 
diffusion of particles with arbitrary size $S$ ($0,1,\dots$) 
gives a dynamical critical exponent $z=\frac{3}{2}$ in the 
KPZ universality class. 

A key ingredient in all the previous 
diffusion processes is that independently of the 
particle's sizes always their order are fixed (even for size 
0 particles). Now we are going to consider more general models 
where this ordering requirement is partially brocken. 
These models will correspond to a generalization of the 
diffusion dynamics of size $S$ particles, for the situation 
where the particles have a negative "size" $S$ 
($S=-1,-2,\ldots$). In those models the particles are ordered 
from left to right and occupy as before the positions 
$\{y(n,t), n =1,\ldots,N_p\}$), with the constraint 
$y(n+1,t) \geq y(n,t) + S$. Then for negative values of $S$ 
the $n$th particle not necessarily is on the right of the  
$(n-1)$th particle, but can be at a distance $-S$ on the left of 
this particle. This negative size $S$ gives a measure of the 
ordering of the particles. In  the limit $S \rightarrow 
-\infty$ we have no ordering at all and the particles should 
behave as noninteracting random walks, with the absense of 
any anomalous behavior. It is interesting to note that the 
related interface model, given by the surface with 
cartesian coordinates $\{n,y(n,t)\}$, can exhibt hills of 
arbitrary height but valleys of maximun depth $-S$, in units 
of lattice  spacing.

	In Fig. 3 we show some examples of our simulations for 
$S=-1$, $p=0.75$,$q=0.25$ and density $\rho=1$ in a lattice 
		of $N_s =150\; 000$ sites.
 As we can see in this figure an anomalous 
behavior is also detected at the critical value $b_c = 0.1667$. 
 This value  
indicate that the conjecture \rf{eq5}, that gives 
$b_c = \frac{1}{6}$ for the present case, is also valid for 
negative values of $S$. Similar results are also obtained 
for other negative values of $S$ (see Table I).

 Inspired by those last results we were able to extend the 
Bethe-ansatz calculations, already known \cite{alcbar1} 
 for particles with size $S \geq 0$, for the  equivalent 
Hamiltonian describing the particles with negative values 
of $S$ \cite{obs2}. 
In the extreme anisotropic limit $p=1$, $q=0$ the eigenenergies of 
the associated Hamiltonian is given by 
\begin{equation} \label{e1}
E = \sum_{j=1}^{N_p} (1 -z_j)/2,
\end{equation}
where the roots $\{z_j\}$ are given by the Bethe-ansatz 
equations
\begin{equation} \label{e2}
(1+z_j)^{N_s-SN_p}(1-z_j)^{N_p} = -2^{N_s} 
e^{-i\frac{2\pi m}{N_p}} \prod_{l=1}^{N_p} 
\frac{z_l-1}{(z_l+1)^{S}},
\end{equation}
where $j=1,\ldots,N_p$, $m= 0,1,\ldots,N_p-1$. The above 
equations are more difficult to handle analitically than 
the corresponding ones in the cases  where $S \geq 0$, 
even at the "half-filling" 
$\chi =\frac{1}{2}$, where a simplification occurs 
\cite{spohn,kim,alcbar1}. The real part of the energy gap 
$G_{N_s} = E_1 - E_0$ behaves for large $N_s$ as $\mbox{Re}
(G_{N_S}) \sim N_s^{-z}$, producing for a pair of lattice sizes 
$N_S,N_S'$ the finite-size 
estimate $z_{N_s,N_s'} = -\ln 
\mbox{Re}(G_{N_s})/\mbox{Re}(G_{N_s'})/\ln (N_s/N_s')$ for 
the dynamical critical exponent $z$. We solved numerically 
\rf{e2} for $S=-1$ and $S=-2$ at the filling $N_p = N_s$. The 
estimates $z_{N_s,N_s+2}$ for some values of $N_s$ are 
presented in Table II. We see clearly that the exponent 
tend toward the KPZ value 
$z = \frac{3}{2}$.

\begin{table}
\caption{\label{t1} 
  Some values of the estimates of the exponent $\nu$ 
of the correlation function \rf{eq3} evaluated at 
$p=0.75$, and for 
molecules of size $S$. These results were obtained at the tag 
velocity $v_{tag}^c = b_c\rho v_p$, with $b_c$ giving by 
(\ref{eq5})}

\begin{ruledtabular}
\begin{tabular}{ccccc}  
  & $S=-2 $ & $S=-1$ & $S= 0$ &  $S= 2$ \\ \hline
 $ \chi = \frac{1}{4}$ & 0.32 & 0.31 & 0.33 & 0.32   \\ 
 $ \chi = \frac{1}{5}$ & 0.31 & 0.32 & 0.31 & 0.33 \\ 
\end{tabular}
\end{ruledtabular}
\end{table}

\begin{table}
\caption{\label{ t1}
 Finite-size estimates $z_{N_s,N_s'}$ of the 
critical dynamical exponent evaluated by solving \rf{e2} for 
$N_s = N_p$.  The extrapolated results are  also shown.}.
\begin{ruledtabular}
\begin{tabular}{cccccc} 
 $N_s,N_s'$ & 50,52  & 100,102 & 150,152 & 200,202 &$\infty$\\ \hline
 $ S=-1$ & 1.61344 & 1.56046 & 1.54122 & 1.53097& 1.500$\pm$0.001 \\  
 $ S=-2$ & 1.72222 & 1.62689 & 1.58882 & 1.56769 & 1.504$\pm$0.005\\  
\end{tabular}
\end{ruledtabular}
\end{table}

In conclusion, our results indicate that as long some type of 
relative ordering exists ($|S| <\infty$), always we are going 
to find anomalous behavior for the sliding-tag correlation 
\rf{eq2}  at the critical tag velocity $v_{tag} = \rho v_p b_c$ 
obtained from \rf{eq1} and \rf{eq5}. This fact can also 
be undertood in terms of the related interface growth model, 
and it implies that as long $|S|$ is finite, we should 
expect the long time fluctuations being governed by a 
KPZ model.

\begin{acknowledgments}
This work was support in part by Conselho Nacional de 
Desenvolvimento Cient\'{\i}fico e Tecnol\'ogico, CNPq, Brazil, 
and FAPESP, S\~ao Paulo, Brazil.
\end{acknowledgments}

%

%parei aqui

% <><><> Final of bibliography <><><><><><>><

\newpage
\Large
\begin{center}
Figure Captions
\end{center}
\normalsize
\vspace{1cm}

\noindent Figure 1 - The time dependence of the sliding-tag 
correlations \rf{eq3} with size $S=2$ particles for some 
values of $b$. The parameters of the Monte Carlo 
simulations are $N_s = 150\; 000$, $N_p =30\; 000$, $p=0.75$. 
%At 
%$b=b_c$ the averaged estimated slope is $2\nu = 0.65 \pm 0.03$.
The averaged estimated slope for the curves are $1.00$ ($b=1$), 
$0.90$ ($b=0.6$), $0.84$ ($b=0.3$) and at $b= b_c$ we have 
the estimate $2\nu = 0.65 \pm 0.03$.

\noindent Figure 2 - Same as figure 1 for size $S=0$ particles.
 The parameters of the Monte Carlo 
simulations are $N_s = 120\; 000$, $N_p =30\; 000$, $p=0.75$. 
The averaged estimated slope for the curves are $0.97$ ($b=1$), 
$0.95$ ($b=0.5$), $0.93$ ($b=0.4$) and at $b= b_c$ we have 
the estimate $2\nu = 0.63 \pm 0.03$.
%At 
%$b=b_c$ the averaged estimated slope is $2\nu = 0.63 \pm 0.03$.

\noindent Figure 3 - Same as figure 1 for size $S=-1$ particles.
 The parameters of the Monte Carlo 
simulations are $N_s = 150\; 000$, $N_p =150\;  000$, $p=0.75$. 
The averaged estimated slope for the curves are $0.86$ ($b=0.4$), 
$0.81$ ($b=0.3$), $0.76$ ($b=0.1$) and at $b= b_c$ we have 
the estimate $2\nu = 0.63 \pm 0.03$.
%At 
%$b=b_c$ the averaged estimated slope is $2\nu = 0.63 \pm 0.03$.


\begin{thebibliography}{99}
%
%
\bibitem{harris} 
T. Harris, J. Appl. Probab. {\bf 2}, 323 (1965).
%
\bibitem{spitzer}
F. Spitzer, Adv. Math. {\bf 5}, 246 (1970).
%
\bibitem{arratia}
R. Arratia, Z. Ann. Probab. {\bf 11}, 362 (1983).
\bibitem{richards}
P. M. Richards, Phys. Rev. B {\bf 16}, 1393 (1977).
\bibitem{fedders}
P. A. Fedders, Phys. Rev. B {\bf 17}, 40 (1978).
\bibitem{alexander}
S. Alexander and P. Pincus, Phys. Rev. B {\bf 18}, 2011 (1978).
\bibitem{beijeren}
H. van Beijeren, K. W. Kehr, and R. Kutner, Phys. Rev. B {\bf 28},
 5711 (1983).

\bibitem{ligget}  T. Ligget, {\it Interacting Particle Systems} 
(Springer-Verlag, Berlin, 1985).
%
\bibitem{demasi}
A. Demasi and P.Ferrari, J. Stat. Phys. {\bf 38}, 603 (1985).
\bibitem{kutner} 
R. Kutner and H. van Beijeren, J. Stat. Phys. {\bf 39}, 3317 (1985).

\bibitem{maba} 
S. N. Majumdar and M. Barma, Phys. Rev. B {\bf 44}, 
5306 (1991).
\bibitem{barma}
M. Barma, J. Phys. A {\bf 25}, L693 (1992).
\bibitem{binder}
P.-M. Binder, M. Paczuski, and M. Barma, Phys. Rev. E {\bf 49}, 1174 
(1994).
%
\bibitem{kpz} M. Kardar, G. Parisi, and Y.C. Zhang, Phys. Rev. Lett.
{\bf 56}, 889 (1986).
%
\bibitem{obs1} In the case where the particles have size $S=0$ 
we have a hole at each lattice point, and consequently the mean 
hole velocity $v_h = 0$, but the critical tag velocity 
is finite and by using the conjecture \rf{eq5} is given by 
$v_{tag}^h = \chi(q-p)(1-\chi)$.
%
\bibitem{alcbar1}
F. C. Alcaraz and R. Z. Bariev, Phys. Rev. E {\bf 60}, 79 (1999).
%
\bibitem{obs2} The Bethe-ansatz equation we obtained are 
similar to  the equation (59) of \cite{alcbar1} with the 
replacements  $N = N_s$, $\tilde{S} = S$, $\epsilon_+ = p$, 
$\epsilon_- = q$,  and $r = n$.
%
%
\bibitem{spohn} L. H. Gwa and H. Spohn, Phys. Rev. Lett. {\bf 68}, 
725 (1992), Phys. Rev. A {\bf 46},
844 (1992).
%
%
\bibitem{kim} D. Kim, Phys. Rev. E {\bf 52}, 3512 (1995).
%
\end{thebibliography}
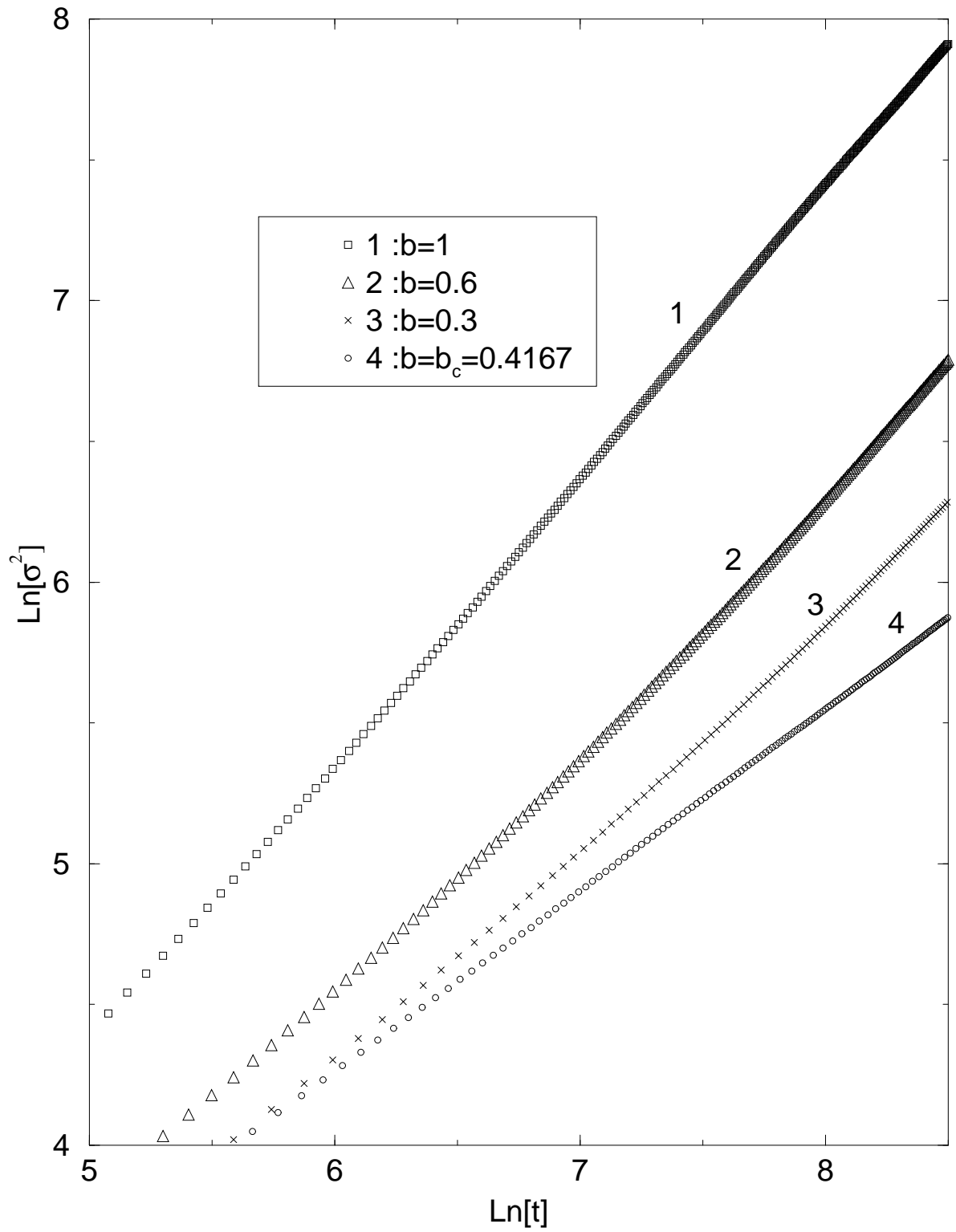
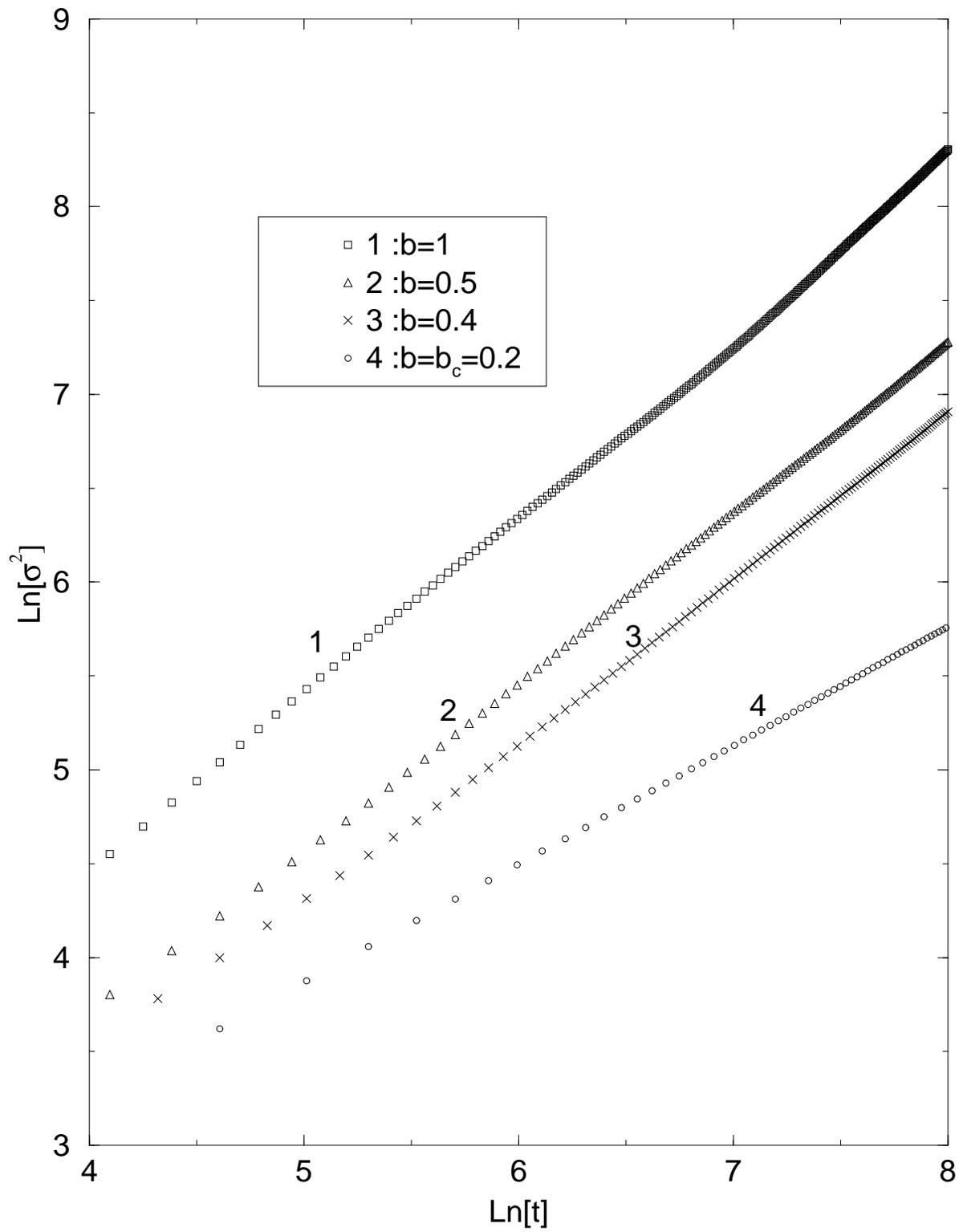
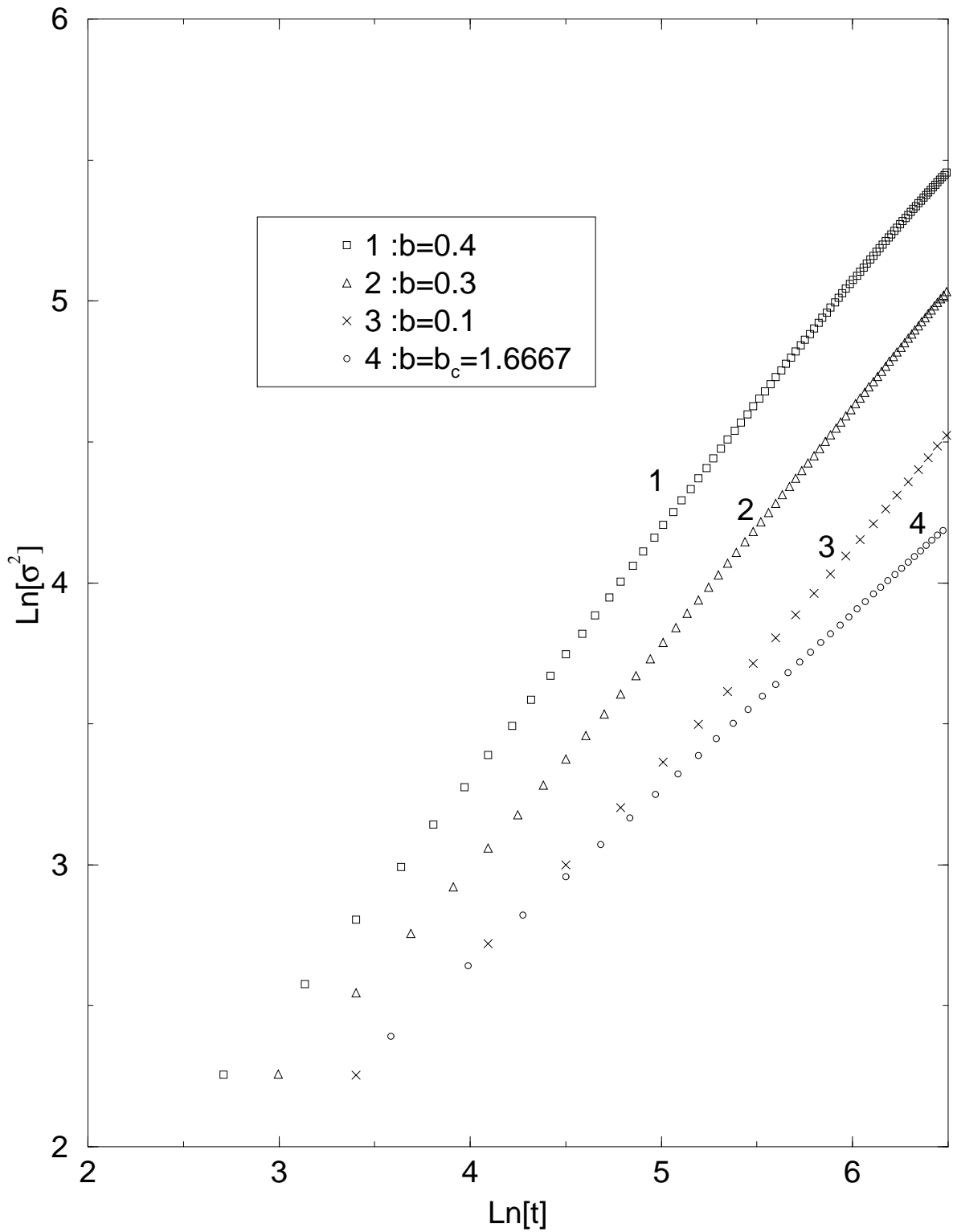
\end{document}